\documentclass[aps,prl,superscriptaddress,notitlepage,twocolumn,longbibliography,10pt]{revtex4-2}

\usepackage{amsmath,amssymb,amsthm,bm}
\usepackage{graphicx}
\usepackage[colorinlistoftodos]{todonotes}
\usepackage[inline]{enumitem}
\definecolor{darkblue}{rgb}{0.1,0.2,0.6}
\definecolor{darkred}{rgb}{0.8,0.1,0.2}
\definecolor{crimson}{RGB}{164,16,52}
\definecolor{darkgreen}{rgb}{0.31,0.62,0.24}
\usepackage[colorlinks=true, allcolors=crimson]{hyperref}
\usepackage{diagbox}
\usepackage{epigraph}
\usepackage{float}
\usepackage{multirow}
\usepackage{tikz}
\usepackage{mathtools}
\usepackage{braket}
\usepackage{soul} 
\usepackage[all]{xy}
\usepackage{lipsum}

\usepackage{bbm}

\newcommand{\Rom}[1]{\uppercase\expandafter{\romannumeral#1}}

\makeatletter
\renewcommand*\env@matrix[1][*\c@MaxMatrixCols c]{%
	\hskip -\arraycolsep
	\let\@ifnextchar\new@ifnextchar
	\array{#1}}
\makeatother

\newtheorem*{claim*}{Claim}

\usepackage[normalem]{ulem}

\definecolor{darkred}{rgb}{0.8,0.1,0.2}

\newcommand*{\ShortSecTitle}[1]{\textit{#1} ---}

\begin{document}
\title{Scar Full Eigenstate Thermalization Hypothesis}

\affiliation{Institute of Theoretical Physics and Department of Physics, University of Science and Technology Beijing, Beijing 100083, China}
\author{Ning Sun}
\affiliation{State Key Laboratory of Surface Physics and Department of Physics, Fudan University, Shanghai 200438, China}
\author{Yanting Cheng}
\email{cyt09240@gmail.com}
\affiliation{Institute of Theoretical Physics and Department of Physics, University of Science and Technology Beijing, Beijing 100083, China}

\begin{abstract}
The eigenstate thermalization hypothesis (ETH) provides a fundamental mechanism for emergent statistical mechanics in isolated chaotic quantum systems, asserting that individual energy eigenstates behave as pseudorandom vectors within an energy window. This enables a complete characterization of nontrivial correlations among matrix elements in the energy eigenbasis, as described by the full ETH ansatz. Nevertheless, this description breaks down in systems exhibiting quantum many-body scars, which host non-thermal eigenstates with extensive energy. In this Letter, we address this problem by formulating the \textit{scar full ETH}, which captures correlations among matrix elements involving scar states. The corresponding scaling forms and factorization properties are established using typicality arguments. Multi-time correlation functions for scar states are then organized in terms of both thermal and scar cumulants, providing a nontrivial reorganization of higher-order correlations. We numerically demonstrate the validity of this framework in the paradigmatic model of quantum scars, the PXP model. Our results pave the way for a systematic understanding of intriguing correlations in systems with quantum many-body scars.
\end{abstract}

\maketitle
\ShortSecTitle{Introduction} The Eigenstate Thermalization Hypothesis (ETH) provides a widely accepted framework for understanding thermalization in isolated quantum many-body systems~\cite{PhysRevA.43.2046,PhysRevE.50.888,Olshanii2008Nature}, bridging quantum chaos and statistical mechanics~\cite{Rigol2016,Foini2019PRE}. It posits that the matrix elements of local observables in the energy eigenbasis take a universal form, in which the diagonal components vary smoothly with energy, while the off-diagonal elements are exponentially suppressed and exhibit random-matrix-like fluctuations. Despite its long history, a complete characterization of these fluctuations has only been developed recently within a refined framework known as the full ETH~\cite{Foini2019PRE,PhysRevLett.129.170603,PhysRevLett.134.140404,PhysRevX.15.011031,Fritzsch:2025ban,Alves:2025jzl,Pathak:2025sys,Vallini:2025vvq,Zhang:2026apc}. Its central ingredients consist of two ansätze. The first ansatz describes the average of products of off-diagonal matrix elements with distinct indices
\begin{eqnarray}\label{eq:relation1}
    \overline{O_{i_1i_2}O_{i_2i_3}\cdots O_{i_qi_1}}=e^{-(q-1)S(E^+)}F^{(q)}(e^+,\vec{\omega}),
\end{eqnarray}
where $O_{i_1 i_2} = \bra{i_1}{\hat{O}}\ket{i_2}$ denotes the matrix element of a local observable $\hat{O}$, and $i_1, \dots, i_q$ label distinct energy eigenstates. $S(E^+)$ denotes the thermal entropy at the average energy $E^+ = (E_{i_1} + \cdots + E_{i_q})/q$, and $\vec{\omega} = (\omega_{i_1 i_2}, \dots, \omega_{i_{q-1} i_q})$ with $\omega_{ij} = E_i - E_j$. $F^{(q)}(e^+,\vec{\omega})$ is a smooth function of the energy density $e^+ = E^+/N$ and $\vec{\omega}$, and is of order unity. Here, $N$ denotes the system size. The statistical average can be understood as an average over eigenstates within a narrow energy window, or over Hamiltonians with small perturbations~\cite{PhysRevLett.129.170603}. The second ansatz states that, when repeated indices appear, the average factorizes in the thermodynamic limit:
\begin{eqnarray}\label{eq:relation2}
    \overline{O_{i_1i_2}\cdots O_{i_ri_1}O_{i_1i_{r+1}}\cdots O_{i_qi_1}}=\nonumber\\
    \overline{O_{i_1i_2}\cdots O_{i_ri_1}}\  \overline{O_{i_1i_{r+1}}\cdots O_{i_qi_1}}.
\end{eqnarray}
Applying these ansätze to multipoint correlation functions leads to a reorganization into fundamental building blocks known as \textit{free cumulants}, a class of connected correlators that probe independence in free probability theory~\cite{speicher1997free}. Numerical demonstrations of the full ETH have been performed in various quantum spin models~\cite{PhysRevLett.134.140404,Alves:2025jzl,Pathak:2025sys,Vallini:2025vvq,Zhang:2026apc}.

\begin{figure}[t]
    \centering
    \includegraphics[width=1.0\linewidth]{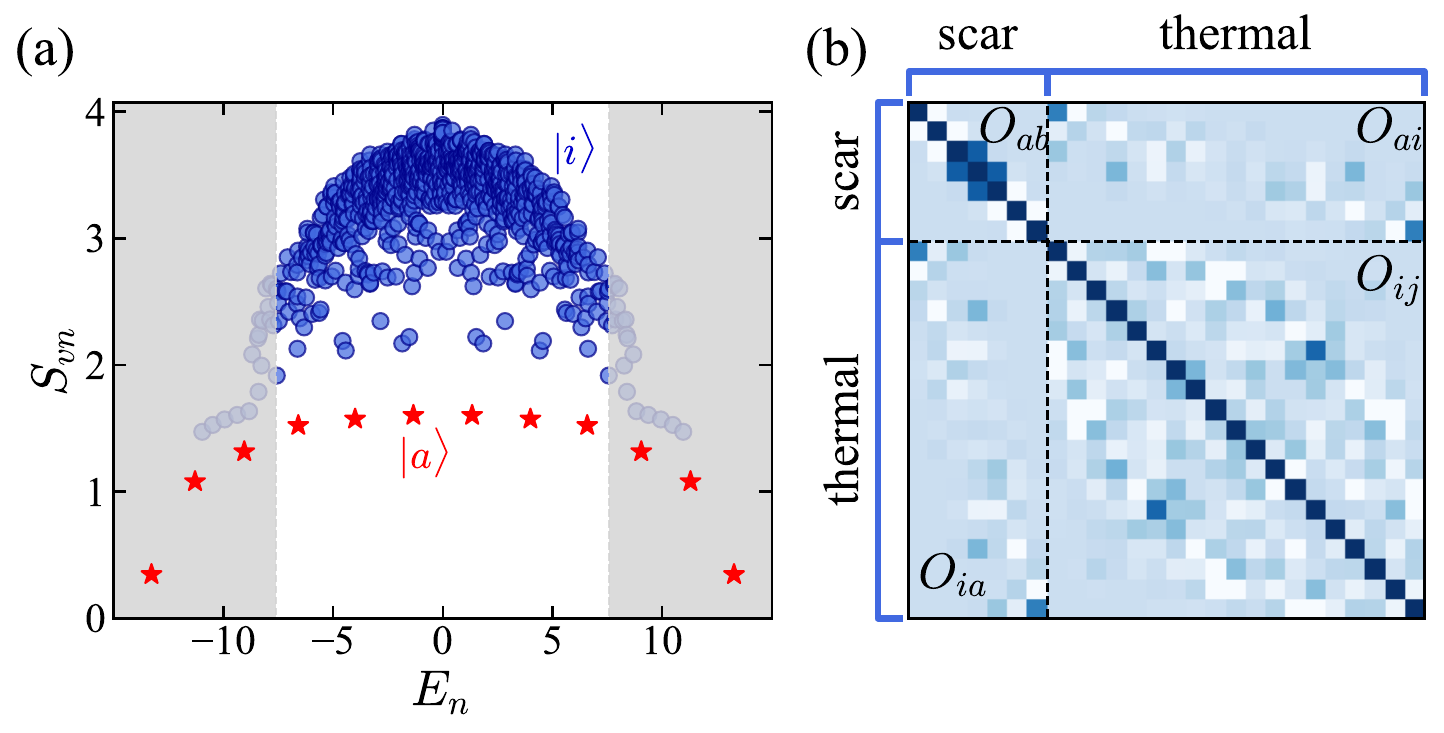}
    \caption{Schematic illustration of the scar full ETH. (a) Typical bipartite entanglement entropy as a function of energy in quantum many-body scar systems. Blue points form the thermal band (labeled by $\ket{i}$), while red stars denote scar states (labeled by $\ket{a}$). The gray regions indicate the many-body band edges, which are not considered in the discussion of ETH. (b) Typical structure of matrix elements in the energy eigenbasis, including the pure scar sector ($O_{ab}$), the thermal sector ($O_{ij}$), and the mixed scar–thermal sectors ($O_{ai}$, $O_{ia}$). We systematically characterize all sectors and establish a complete scar full ETH framework. } 
    \label{fig:schematic}
\end{figure}

Despite the success of the full ETH in fully chaotic systems, a fundamental challenge arises when extending its scope to weakly ergodicity-breaking systems~\cite{Papic2018NP,Papic2018PRB,AKLT2018PRB,Bernevig2018AKLT,Choi2019PRL,Lin2019PRL,Papic2021NP,Bernevig2022review}. In particular, in systems with quantum many-body scars, the majority of states (labeled by $\ket{i}$) remain thermal, while nonthermal scar states (labeled by $\ket{a}$)  with low entanglement emerge even in the center of the many-body spectrum (see Fig.~\ref{fig:schematic}(a)). Although these scar states occupy only a vanishing fraction of the Hilbert space in the thermodynamic limit, they give rise to striking quantum dynamics characterized by persistent temporal oscillations, as observed in Rydberg atom arrays~\cite{Lukin2017Nature,Lukin2021Science}. On the other hand, a naive application of the traditional full ETH, which assumes complete thermalization throughout the entire spectrum, would entirely overlook the intrinsic dynamical effects associated with scar states, rendering it incapable of faithfully describing scarred quantum many-body systems. This naturally raises the following question: Is it possible to generalize the full ETH ansatz to characterize weakly ergodicity-breaking systems hosting quantum many-body scars?

In this Letter, we address this question by formulating the scar full ETH (SFETH) ansätze, which capture the enriched correlations between matrix elements involving scar states. We establish the corresponding scaling forms and their factorization properties in the thermodynamic limit using typicality arguments. Applying these ansätze to multipoint functions on scar states, we demonstrate that higher-order correlations are organized not only by thermal free cumulants, but also by a new class of \emph{scar free cumulants}, which encode the hybrid structure of thermal–scar contributions. We further substantiate our theory through numerical simulations of the paradigmatic PXP model~\cite{RevModPhys.82.2313,PhysRevLett.106.025301,PhysRevB.99.161101,Papic2018NP,Papic2018PRB,Papic2021NP,Ho2019PRL,Lin2019PRL}. Our results provide a unified framework for understanding higher-order correlations and information dynamics in systems with quantum many-body scars.

\ShortSecTitle{SFETH ansätze} 
In systems with quantum many-body scars, the matrix elements of an observable $\hat{O}$ in the energy eigenbasis can be classified into three sectors, as illustrated in Fig.~\ref{fig:schematic}(b). The matrix elements in the thermal sector, $O_{ij}$, do not depend on the presence of scar states. Therefore, their intra-sector correlations exhibit a structure similar to that of fully chaotic systems, as described by the traditional full ETH ansätze \eqref{eq:relation1} and \eqref{eq:relation2}. In contrast, the matrix elements in the pure scar sector, $O_{ab}$, depend entirely on the details of the scar states, which are not constrained by typicality arguments. Fortunately, the number of such elements scales only quadratically with the number of scar states, which itself typically grows polynomially with system size. Finally, the matrix elements in the mixed scar-thermal sector $(O_{ai}, O_{ia})$ encode information about both thermal and scar states. They exhibit novel correlations with matrix elements in the thermal sector, which constitute the main focus of our SFETH ansätze. 

The first ansatz of the SFETH describes the the average of such scar-involving high-order matrix elements products
\begin{equation}\label{average}
    \overline{O_{ai_1}O_{i_1i_2}\cdots O_{i_{q-1}i_q}O_{i_qb}}=e^{-qS(E_{\mathrm{th}})}P^{(q+1)}_{ab}(\vec{\omega}).
\end{equation}
Here, we assume that the indices corresponding to thermal states are distinct. $E_{\mathrm{th}} = (E_{i_1} + \cdots + E_{i_q})/q$ is the central thermal energy, and $e_{\mathrm{th}} = E_{\mathrm{th}}/N$ is the energy density. For conciseness, we have introduced the notation $\vec{\omega} = (\omega_{ai_1}, \omega_{i_1 i_2}, \dots, \omega_{i_{q-1} i_q}, \omega_{i_qb})$. $P^{(q+1)}_{ab}(\vec{\omega})$ is a smooth function of order unity that decays to zero for large $\omega$, since the local observable $O$ can only change the energy of the system locally. The second ansatz states that, for products with repeated indices, the average factorizes in the thermodynamic limit:
\begin{equation}\label{factorization}
\begin{aligned}
    \overline{O_{ai_1}\cdots O_{i_{r-1}i_r}O_{i_ri_{r+1}}\cdots O_{i_si_r}O_{i_ri_{s+1}}\cdots O_{i_q b}}\\
    =\overline{O_{ai_1}\cdots O_{i_{r-1}i_r}O_{i_ri_{s+1}}\cdots O_{i_q b}}\ \overline{O_{i_ri_{r+1}}\cdots O_{i_si_r}},
\end{aligned}
\end{equation}
where the last factor in the second line can be determined using the traditional full ETH ansatz \eqref{eq:relation1}. 

These SFETH ansätze are inspired by typicality arguments. The central assumption is that the thermal energy eigenstates behave as orthogonal Haar-random states. We then evaluate the Haar-random average for the product of matrix elements as
\begin{eqnarray}
\overline{O_{a i_1}\cdots O_{i_q b}} =
\int_{\mathrm{Haar}} d\hat{U} \bra{a}\hat{O}\hat{U}\ket{i_1}\cdots \bra{i_q}\hat{U}^\dagger \hat{O}\ket{b},
\end{eqnarray}
where $\hat{U}$ is sampled from the unitary group with the Haar measure. By explicitly performing the integration over $\hat{U}$ using Weingarten calculus~\cite{collins2022weingarten}, as detailed in the Supplementary Material~\cite{SM}, the result exhibits the expected scaling in Eq.~\eqref{average}, where the thermodynamic entropy is identified as $S=\ln D$, with $D$ denoting the dimension of the unitary matrix $\hat{U}$. Furthermore, computing similar average with repeated indices leads to the factorization property in Eq.~\eqref{factorization}. 

Alternatively, the factorization property can be justified directly from the average-value relations in Eqs.~\eqref{eq:relation1} and \eqref{average}. By counting the number of matrix elements, one finds that the right-hand side of Eq.~\eqref{factorization} scales as $e^{-qS}$. On the other hand, the correction to Eq.~\eqref{factorization} originates from fluctuations of $O_{ai_1}\cdots O_{i_{r-1}i_r}O_{i_ri_{s+1}}\cdots O_{i_q b}$ and $O_{i_ri_{r+1}}\cdots O_{i_si_r}$ about their expectation values. Assuming that these fluctuations scale in the same way as products of matrix elements with no repeated indices~\cite{PhysRevE.99.042139}, one may estimate the contribution from the correlation between $O_{ai_1}\cdots O_{i_{r-1}i_r}O_{i_r'i_{s+1}}\cdots O_{i_q b}$ and $O_{i_ri_{r+1}}\cdots O_{i_si_r'}$, where $i_r'$ is an independent index. According to Eq.~\eqref{average}, this correlation scales as $e^{-(q+1)S}$. The correction is therefore exponentially suppressed by an additional factor of $e^{-S}$ compared with the leading contribution, and can thus be neglected in the thermodynamic limit.

\begin{figure}
    \centering
    \includegraphics[width=1.0\linewidth]{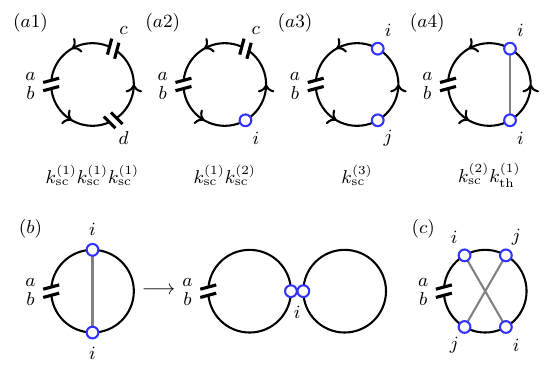}
    \caption{(a) Diagrammatic representation of the decomposition of correlation functions in the scar state. (b) Factorization of the scar free cumulants in Eq.~(\ref{factorization}) for $q=3$, where the expression factorizes into a second-order scar free cumulant and a first-order thermal free cumulant. (c) Example of a crossing diagram for $q=5$, which vanishes in the thermodynamic limit. In all panels, the solid lines represent insertions of the operator $\hat{O}$. Each circular node and double line represent the insertion of projection operators onto thermal states and scar states, respectively. 
    Gray lines denote identical thermal indices.
    }  
    \label{fig:diagram}
\end{figure}

\ShortSecTitle{Scar free cumulants} 
Next, we apply the SFETH ansätze to correlation functions. This enables the reduction of multipoint correlation functions to free cumulants in systems with quantum many-body scars, analogous to the role played by the full ETH in thermal systems. In this case, however, the structure is enriched: the fundamental building blocks consist of both thermal and scar free cumulants. In addition, the calculation provides a direct relation between the function $P_{ab}^{(q)}(\vec{\omega})$ and correlation functions on scar states in the frequency domain.

To begin with, we investigate multipoint thermal correlation functions, which are defined as\begin{eqnarray}
\mathcal{F}^{(q)}_\beta(\vec{t})=\langle \hat{O}(t_1)\hat{O}(t_2)\cdots \hat{O}(t_{q-1})\hat{O}(0)\rangle_\beta,
\label{Green's_function}
\end{eqnarray}
where $\langle \cdot \rangle_\beta = \mathrm{Tr}\left( e^{-\beta \hat{H}} \cdot \right)/Z$ with the partition function $Z=\mathrm{Tr}(e^{-\beta \hat{H}})$, and we have introduced the notation $\vec{t}=(t_1,\cdots,t_{q-1})$. By inserting a complete set of energy eigenstates between operators, the evaluation reduces to summing over all possible contraction structures. In systems without quantum scars, the application of Eq.~\eqref{eq:relation1} and \eqref{eq:relation2} reorganizes the correlation function in terms of thermal free cumulants~\cite{PhysRevLett.129.170603}
\begin{equation}\label{eq:thermal_cumulant}
\begin{aligned}
k_{\mathrm{th},\beta}^{(q)}(\vec{t}) &\equiv \frac{1}{Z} \sum_{i_1 \neq \cdots \neq i_q} e^{-\beta E_{i_1}} O(t_1)_{i_1 i_2} \cdots O(0)_{i_q i_1},\\
&=\int d\vec{\omega}~e^{i\vec{\omega}\cdot\vec{t}-\beta\vec{\omega}\cdot\vec{l}_q}F^{(q)}(e_\beta,\vec{\omega}).
\end{aligned}
\end{equation}
Here, $\vec{l}_q=\big((q-1)/q,(q-2)/q,\cdots, 1/q\big)$, and the energy density $e_\beta$ is determined by the thermodynamic relation. Now, we analyze the contribution of scar states to thermal correlation functions. For example, if a single scar state is inserted in place of $i_2$, the corresponding contribution to $\mathcal{F}^{(q)}_\beta(\vec{t})$ is given by $$\frac{1}{Z} \sum_{i_1 \neq i_3\neq\cdots \neq i_q,a} e^{-\beta E_{i_1}} O(t_1)_{i_1 a} O(t_2)_{ai_3}\cdots O(0)_{i_q i_1}.$$ Using the average of the matrix elements involving the scar state in Eq.~\eqref{average}, the matrix elements scale as $e^{-(q-1)S}$, which is compensated by the summation over the remaining thermal indices. As a consequence, this contribution scales as $Z^{-1}$ and is therefore exponentially suppressed in the thermodynamic limit. Consequently, the multipoint thermal correlation functions do not receive contributions from the scar states, and the traditional expansion in terms of thermal free cumulants remains valid.

Nevertheless, the scar states do exhibit novel dynamical signatures that are not captured by thermal correlation functions. These signatures can be detected by evaluating scar correlation functions defined as
\begin{equation}
\mathcal{F}^{(q)}_{ab}(\vec{t})=\bra{a}\hat{O}(t_1)\hat{O}(t_2)\cdots \hat{O}(t_q)\ket{b}.
\end{equation}
Here, $\vec{t}=(t_1,\cdots,t_{q})$. The explicit dependence on scar states necessitates the application of SFETH to decompose these scar correlation functions. Similar to the thermal correlators, we insert complete sets of energy eigenstates between the operators and separate the contributions from scar states and thermal states at each insertion. As an example, we consider the calculation for $q=3$. A diagrammatic representation of the different contributions is illustrated in FIG.~\ref{fig:diagram}(a1–a4), where the insertion of thermal states is represented by circular nodes, and the insertion of scar states is represented by double lines. In panels (a1–a3), the double lines partition the diagram into distinct pieces, each containing thermal states with different indices. To describe these contributions, we introduce the \textit{scar free cumulants} 
\begin{equation}
\begin{aligned}
k_{\mathrm{sc},ab}^{(q)}(\vec{t}) &\equiv \sum_{i_1 \neq \cdots \neq i_{q-1}}
O(t_1)_{a i_1} \cdots O(t_{q})_{i_{q-1} b},\\
&=\int d\vec{\omega}~e^{-i\vec{\omega}\cdot \vec{t}}P^{(q)}_{ab}(\vec{\omega}),
\end{aligned}
\end{equation}
which is the Fourier transform of the function $P^{(q)}_{ab}(\vec{\omega})$. Then, the contribution from (a1-a3) becomes 
\begin{equation}\label{eq:1-3}
\begin{aligned}
&\text{(a1-a3)}=k_{\mathrm{sc},ac}^{(1)}(t_1)k_{\mathrm{sc},cd}^{(1)}(t_2)k_{\mathrm{sc},db}^{(1)}(t_3)\\
&+k_{\mathrm{sc},ac}^{(2)}(t_1,t_2)k_{\mathrm{sc},cb}^{(1)}(t_3)+k_{\mathrm{sc},ac}^{(1)}(t_1)k_{\mathrm{sc},cb}^{(2)}(t_2,t_3)\\
&+k_{\mathrm{sc},ab}^{(3)}(t_1,t_2,t_3).
\end{aligned}
\end{equation}
Here, the summations over the repeated indices $c$ and $d$ are implicit. The remaining task is to evaluate the contribution from panel (a4). Using the factorization relation, we find
\begin{equation}\label{eq:a4con}
\text{(a4)}=\sum_{i}\overline{O(t_1)_{ai}O_{ii}O(t_3)_{ib}}=\sum_{i}\overline{O(t_1)_{ai}O(t_3)_{ib}} \times\overline{O}_{ii}.
\end{equation}
As mentioned above, since the operator $\hat{O}$ is local, the energy difference between $i$ and $a$ (or $b$) is of order unity. This is guaranteed by the decay of the function $P_{ab}^{(2)}(\vec{\omega})$ with increasing frequency $\vec{\omega}$. Therefore, the energy density of the state $i$ is identical to that of the state $a$ (or $b$) in the thermodynamic limit. Since the thermal matrix elements depend only on the average energy density, according to Eq.~\eqref{eq:relation1}, we have
\begin{equation}\label{eq:4}
\text{(a4)}=k_{\mathrm{sc},ab}^{(2)}(t_1,t_3)\times k_{\mathrm{th},\beta_{ab}}^{(1)}.
\end{equation}
Here, $\beta_{ab}$ is determined by matching the energy density $e_{\beta_{ab}}=(E_a+E_b)/2N$~\footnote{We choose a form that is symmetric in $a$ and $b$. In the thermodynamic limit, one may equivalently use the energy density $E_a/N$ or $E_b/N$.}. Summing the contributions in Eqs.~\eqref{eq:1-3} and \eqref{eq:4} leads to a reorganization of the three-point scar correlation function $\mathcal{F}^{(3)}_{ab}(\vec{t})$ in terms of both thermal and scar free cumulants.

This procedure naturally generalizes to higher-order scar correlation functions. After inserting complete sets of energy eigenstates, we first classify the diagrams according to the insertions of scar states. Scar insertions cut the diagram into ``disjoint'' pieces (arcs), each containing possible insertions of thermal states. We then separate the contributions according to whether the thermal indices within each segment coincide, as exemplified in FIG.~\ref{fig:diagram}(a3) and (a4). Each segment can then be classified in a manner similar to that for thermal diagrams~\cite{PhysRevLett.129.170603}: (1) Arc diagrams, in which all nodes on every arc are distinct; (2) Cactus diagrams, in which trees of thermal loops are attached to the arcs; and (3) Crossing diagrams, which are neither arc nor cactus diagrams. An example of a crossing diagram for $q=5$ is shown in FIG.~\ref{fig:diagram}(c). Using \eqref{eq:relation1} and \eqref{average}, its contribution scales as
\begin{equation}
\sum_{ij}{O_{a i} O_{i j}  O_{j b}}\times {|O_{i j}|^2}\sim e^{-S}.
\end{equation}
Hence, crossing diagrams are parametrically suppressed by the inverse density of states and can be neglected in the thermodynamic limit \cite{SM}. Therefore, only arc diagrams and cactus diagrams contribute, which leads to the general decomposition using the scar free cumulants together with thermal free cumulants:
\begin{equation}\label{eq:res}
\mathcal{F}^{(q)}_{ab}(\vec{t})=
\sum_{\pi\in \Pi(q)}
\bigg(\prod_{q'}k_{\mathrm{sc}}^{(q')}(\vec{t}_{q'})\bigg)_{ab} \prod_{q''}k_{\mathrm{th},\beta_{ab}}^{(q'')}(\vec{t}_{q''}) . 
\end{equation}
Here, $\pi$ denotes one of all the legal \textit{non-crossing} partitions $\Pi(q)$, satisfying $\sum {q'} + \sum {q''} = q$, and the time vectors $\vec{t}_{q'}$ and $\vec{t}_{q''}$ are uniquely determined from the partition. For conciseness, we view scar free cumulants as matrices in the scar space and use matrix multiplication to define their products. This result further provides a definition of the function $P_{ab}^{(q)}(\vec{\omega})$ introduced in the SFETH ansatz \eqref{eq:relation1}. Since the $q$-point correlation function $\mathcal{F}^{(q)}_{ab}(\vec{t})$ in Eq.~\eqref{eq:res} contains only scar free cumulants with $q'\leq q$, we can iteratively express $k_{\mathrm{sc},ab}^{(q)}(\vec{t}{q})$ in terms of correlation functions $\mathcal{F}^{(q')}_{ab}(\vec{t})$ with $q'\leq q$. Performing an inverse Fourier transform then yields an expression for $P_{ab}^{(q)}(\vec{\omega})$. 

\ShortSecTitle{Numerical verification} 
Finally, we provide a numerical demonstration of our SFETH theory in the spin-$1/2$ PXP model~\cite{PhysRevLett.106.025301,PhysRevB.99.161101,Papic2018NP,Papic2018PRB,Papic2021NP,Ho2019PRL,Lin2019PRL}, which has been widely realized experimentally in Rydberg atom arrays~\cite{Lukin2017Nature,Lukin2021Science,Labuhn2016Nature}. The Hamiltonian of the PXP model is given by
\begin{eqnarray}
\hat{H}_{\mathrm{PXP}} = \sum_{j=1}^N\hat{P}_{j-1}\hat{\sigma}^x_j\hat{P}_{j+1},
\end{eqnarray}
where $j$ labels the lattice sites arranged on a one-dimensional lattice with periodic boundary conditions. Here, $\hat{P}_j = \frac{1-\hat{\sigma}^z_j}{2}$ is the projector onto the $\ket{\downarrow}$ state, originating from the Rydberg blockade effect~\cite{PhysRevLett.85.2208,PhysRevLett.87.037901,RevModPhys.82.2313,Urban2009NP,CNRS2009NP}, which enforces that no two neighboring spins can simultaneously occupy the $\ket{\uparrow}$ state. It has been established that the PXP model hosts a set of quantum many-body scar states that violate the conventional ETH description~\cite{Papic2018NP,Papic2018PRB,Choi2019PRL,Lin2019PRL,Papic2021NP,Bernevig2022review}.

\begin{figure}
    \centering
    \includegraphics[width=1.0\linewidth]{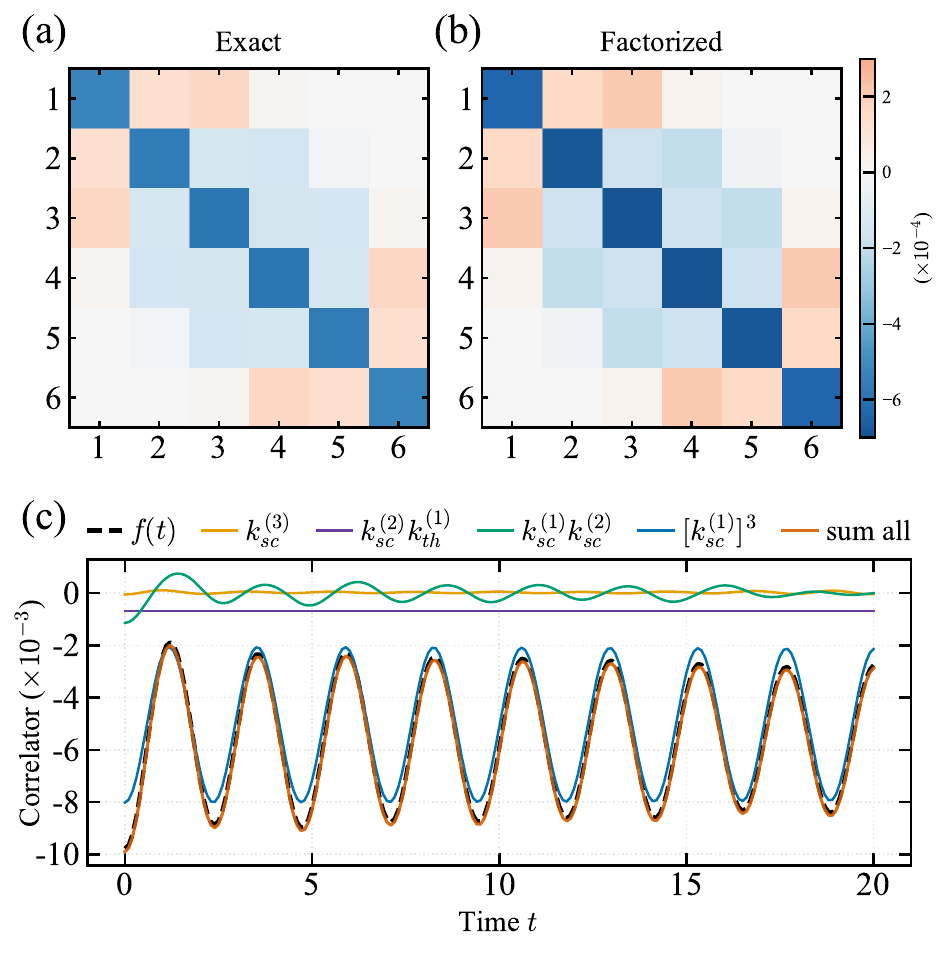}
    \caption{Numerical verification of the SFETH in the PXP model with $N=22$ and local operator $\hat{O}=\frac{1}{2N}\sum_{j=1}^N \hat{\sigma}_j^z$. The scar states are obtained in the $k=0$ even-parity sector with periodic boundary conditions. (a,b) Test of the SFETH factorization \eqref{eq:factorization} for six scar states near the center of the spectrum, where (a) shows the exact result and (b) shows the factorized result. (c) Comparison between the multi-time correlation function and its SFETH free-cumulant decomposition. Here, we consider the three-point function $\mathcal{F}_{aa}^{(3)}(0,t,0)$, where $\ket{a}$ is the scar state closest to the center of the spectrum.}
    \label{fig:simulation}
\end{figure}

We perform exact diagonalization for a system of size $N=22$ by restricting to the subspace with total momentum $k=0$ and even parity under reflection. To restrict our discussion to this subspace, we choose the following translationally invariant local observable $\hat{O}=\frac{1}{2N}\sum_{j=1}^N\hat{\sigma}_i^z$. We first test the factorization of the contributions to the three-point correlation function on scar states in Eqs.~\eqref{eq:a4con} and \eqref{eq:4}. For simplicity, we set $\vec{t}=0$, and the relation to be tested becomes
\begin{equation}\label{eq:factorization}
\sum_i O_{a i} O_{i i} O_{i b} = \Big(\sum_i O_{a i} O_{i b}\Big) \cdot \Bigg(\frac{1}{Z} \sum_j O_{j j} e^{-\beta_{ab} E_j}\Bigg),
\end{equation}
Here, the inverse temperature $\beta_{ab}$ can be determined numerically. In FIG.~\ref{fig:simulation}(a) and (b), we present density plots of the \textit{l.h.s.} and \textit{r.h.s.}, respectively. We choose six scar states from the middle band, corresponding to the unshaded region in Fig.~\ref{fig:schematic}(a), where the bulk thermal states are sufficiently thermalized for the SFETH description to be expected to apply. Different pixels in the density plots correspond to different pairs of initial and final scar states. The excellent agreement between the two panels demonstrates the validity of the factorization predicted by the SFETH ansätze. We also provide numerical evidence for the negligible contribution of crossing diagrams in the Supplementary Material~\cite{SM}.

We further demonstrate the decomposition of multipoint correlation functions on scar states as a function of time. As an example, we consider the three-point function $\mathcal{F}_{aa}^{(3)}(0,t,0)$, which is related to the experimental observable in \cite{You2025PRL,Xiang:2024cdw}. Here, we choose the scar state closest to the center of the spectrum, indicated in Fig.~\ref{fig:schematic}(a), as both the initial and final scar state $\ket{a}$. We compare the exact correlation function with its decomposition in Eqs.~\eqref{eq:1-3} and \eqref{eq:4}. The result is shown in FIG.~\ref{fig:simulation}(c). The colored curves represent the four different types of contributions defined in the SFETH decomposition, while the red curve denotes their sum. The excellent agreement between the exact result, shown as the black dashed line, and the red curve demonstrates that the SFETH accurately captures higher-order scar--scar correlations in systems with quantum many-body scars.

\ShortSecTitle{Discussions}
In this work, we propose an extended scar full ETH framework to systematically characterize higher-order ETH structures in systems with quantum many-body scars. Based on the SFETH ansätze, which describes the scaling behavior and factorization properties of scar-involving higher-order products of matrix elements, we introduce the concept of \textit{scar free cumulants}. Together with the conventional thermal free cumulants, they constitute the fundamental building blocks governing multipoint correlation functions defined on scar states, which exhibit the celebrated persistent oscillations. Our theoretical predictions are explicitly demonstrated through numerical simulations of the PXP model, a paradigmatic model of quantum many-body scars. Our results establish a unified framework for understanding higher-order correlations and nonthermal dynamical behavior in scarred quantum systems.

We conclude this work with a few remarks. First, it would be worthwhile to investigate whether the extended SFETH framework can be connected to effective descriptions of scar dynamics, such as quasiparticle or collective-mode pictures underlying persistent oscillations~\cite{Lukin2017Nature,Choi2019PRL,Ho2019PRL,Papic2020PRX,Papic2021NP}. Second, it would be interesting to generalize the present discussion to other systems with weak ergodicity breaking, such as those exhibiting Hilbert-space fragmentation~\cite{Pollman2020PRX,Bernevig2022review,Serbyn2023SCIP}. In these systems, the Hilbert space is separated into disconnected blocks, while the eigenstates within each block can still remain thermal. Extending our framework to such systems may lead to distinct scaling forms for intra-block and inter-block matrix elements. Finally, it would be particularly interesting to understand whether the analytical properties of the function $P^{(q)}_{ab}(\vec{\omega})$ impose constraints on the rates of spin relaxation or information scrambling in many-body systems, analogous to the derivation of the chaos bound in thermal systems~\cite{PhysRevLett.123.230606}. We leave these questions for future work.

\ShortSecTitle{Acknowledgments}
We thank Pengfei Zhang, Shang Liu, and Lei Feng for the helpful discussion. Y. C. is supported by NSFC Grant No. 12374251. 
N. S. acknowledge the Scientific Research Innovation Capability Support Project for Young Faculty (Grant No. ZYGXQNJSKYCXNLZCXM-I14).

\bibliography{Bib_Refs.bib}
\end{document}


\title{Supplementary Information for ``Scar Full Eigenstate Thermalization Hypothesis"}

\affiliation{Institute of Theoretical Physics and Department of Physics, University of Science and Technology Beijing, Beijing 100083, China}
\author{Ning Sun}
\affiliation{State Key Laboratory of Surface Physics and Department of Physics, Fudan University, Shanghai 200438, China}
\author{Yanting Cheng}
\email{cyt09240@gmail.com}
\affiliation{Institute of Theoretical Physics and Department of Physics, University of Science and Technology Beijing, Beijing 100083, China}

\date{\today}

\begin{abstract}
In this supplementary material, we present results for (1) details of three-point out-of-time-order scar-scar correlation functions, (2) diagrammatic decomposition of four-point functions, (3) numerical verification of vanishing crossing diagrams, and (4) review of Weingarten calculus and examples of SFETH scaling form from typicality arguments. 
\end{abstract}

\maketitle

\subsection{Supplementary Note 1. Three-point out-of-time-order scar-scar correlation functions }

Take $\bra{a}\hat O(0)\hat O(t)\hat O(0) \ket{b}$ as an example, where $\hat O=\frac{1}{2N}\sum_i^N\hat\sigma_i^z$ and $\hat O(t)$ is the operator in Heisenberg picture.
With the help of diagrams in Fig.~2(a1-a4) and Eqs.~(10) and (12), we have
\begin{equation}\label{eqn:Sz0SztSz0-lhs}
\hspace{-10ex}
\begin{aligned}
&\bra{a} \hat O(0) \hat O(t) \hat O(0) \ket{b} 
= \bra{a}\hat Oe^{iHt}\hat Oe^{-iHt}\hat O \ket{b} 
= \\
&= 

\end{aligned}
\end{equation}

Here, each term means:
\begin{equation}\label{term1}
(1): \quad 
k_{\mathrm{sc},ab}^{(3)}(t) = 
%
=O_{ac}O_{cd}O_{db}e^{i(E_c-E_d)t}. 
\end{equation}

To summary, the SFETH predicts the $\bra{a}O(0)O(t)O(0) \ket{b}$ should be equal to
\begin{equation}\label{eqn:Sz0SztSz0-rhs}
(1) + (2) + \sum_c ((3c) + (3'c)) + \sum_{c,d}(4cd). 
\end{equation}
In the maintext Fig.~3(c), we have numerically checked each term above and verified that \eqref{eqn:Sz0SztSz0-rhs} equals to the \textit{l.h.s.} of \eqref{eqn:Sz0SztSz0-lhs} to good agreement numerically.

\subsection{Supplementary Note 2. Diagrammatic representation of four-point function decomposition }

A diagrammatic representation of the fully decomposition of the four-point scar-relevant correlation (with non-repeated thermal indices) functions within the SFETH framework is provided here. The time direction is omitted, without loss of generality. 
(And dummy thermal indices in the factorized diagrams are also omitted for conciseness.) 
\begin{equation}
\hspace{-12ex}
\begin{aligned}
&\quad \sum_{c,d,f}
 
\  . 
\end{aligned}
\end{equation}

\subsection{Supplementary Note 3. Numerical testing of vanishing crossing terms }
\begin{figure}
    \centering
    \includegraphics[width=0.4\linewidth]{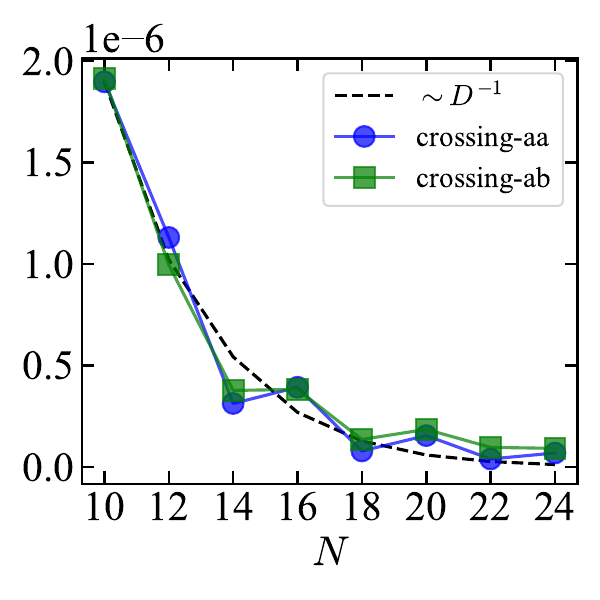}
    \caption{Numerical results of the crossing term in the PXP model. Blue dots and line denotes the crossing term with the same initial and final states both the scar state $\ket{a}$ depict from the middlest of the energy spectrum. Green dots and line denotes the crossing term with initial state $\ket{a}$ and final state $\ket{b}$, which is the scar state right next to $\ket{a}$ in the energy spectrum. Dashed line marks an inverse of the Hilbert space dimension scaling. }
    \label{fig:crossing}
\end{figure}

The lowest order crossing term in the SFETH framework appears in the five-point scar-scar correlation functions, which is 
\begin{equation}
\sum_{i\neq j}O_{ai}O_{ij}O_{ji}O_{ij}O_{jb}\sim \mathcal{O}(D^{-\alpha}),
\end{equation}
or diagrammatically, 
\begin{equation}
%
 
\sim\mathcal{O}(D^{-\alpha})\longrightarrow 0. 
\end{equation}
Here since $\hat O$ is some local operator. The contribution of such crossing diagrams is subleading in $\log D$, as we now argue (similar to \cite{fulleth_prl129,fulleth_prl134}). 
\begin{equation}
\begin{aligned}
\sum_{i\neq j}\overline{O_{ai}O_{ij}O_{jb}}\ \overline{|O_{ij}|^2}
&= \int dE_i dE_j e^{S(E_i)}e^{S(E_j)} e^{-2S(E^+)} e^{-S(E^+)} f_{ab}(\omega) \\
&\simeq \int dEd\omega e^{-S(E)}f(\omega) \sim \mathcal{O}(e^{-S})
\sim \mathcal{O}(D^{-\alpha}). 
\end{aligned}
\end{equation}
We numerically test this term with PXP model and local operator $\hat O=\frac{1}{2N}\sum_i^N\hat\sigma_i^z$. 
With increasing system size $N$, the crossing term decays rapidly as the inverse of hilbert space dimension.
In Fig.~\ref{fig:crossing}, we show both the $a$-to-$a$ crossing term and $a$-to-$b$ crossing term, in blue and green dots and lines, respectively, and compare them with an inverse Hilbert space dimension scaling $D^{-1}$. 
Here $a$ denotes the middlest scar state and $b$ denotes the scar state right next to $a$ in energy spectrum. 
It shows that, with increasing system size $N$, both the $a$-to-$a$ and $a$-to-$b$ crossing terms decays rapidly as approximately the scaling of $\sim D^{-1}$, suggesting that in the thermodynamic limit the crossing terms indeed vanishes and contributes zero to the physical scar correlation functions.  

\subsection{Supplementary Note 4. Scaling ansatz of SFETH from typicality arguments }
In the maintext, we formulate the SFETH scaling ansatz from typicality arguments, which assumes the bulk thermal states are Haar random states that can be averaged in a narrow energy window. 
So in this section, we first briefly review the Weingarten calculus regarding Haar random states and Haar random average, and then use it to give a few simplest examples of the scaling form of SFETH based on the typicality arguments. 

The defining property of Haar measure is
\begin{equation}\label{eqn:Haar_def}
    \int_{Haar}dUf(U) = \int_{Haar}dUf(UW) = \int_{Haar}dUf(WU),
\end{equation}
satisfied for any $f$, as a function of $U$,
\footnote{To be nontrivial, $f$ should not be only function of $U$ but also $U^\dag$, $f(U,U^\dag)$. }
and any arbitrary $W$,
with the normalization
\begin{equation}\label{eqn:norm}
    \int_{Haar}dU = 1. 
\end{equation}

\paragraph{First moment. }
Since \eqref{eqn:Haar_def} should be satisfied for any arbitrary $f(U)$, we firstly consider to choose $f$ as the first moment, $f(U) = U\otimes U^\dag$. 
Then we have,
\begin{equation}
    \tilde{I} = \int_{Haar}dUU\otimes U^\dag = \int_{Haar}dU (WU)\otimes (U^\dag W^\dag) = \int_{Haar}dU(UW)\otimes (W^\dag U^\dag). 
\end{equation}
By Shur's Lemma, and a few steps of simplification, we obtain for the 1st moment,
\begin{equation}\label{eqn:Haar_1st_moment}
\int_{Haar}dU \  

\end{equation}

\paragraph{Second moment.} 
Similarly for the second moment, we have,
\begin{equation}\label{eqn:Haar_2nd_moment}
\int_{Haar}dU \  
%
. 
\end{equation}
Here $d$ is the Hilbert space dimension, exponential in system size $N$, i.e., $d\sim \exp(N)$. 

\paragraph{General $n$th moment.}
Similarly for higher moments. Generally we have,
\begin{equation}\label{eqn:Haar_nth_moment}
\int_{Haar}dU 
%
\end{equation}
where $\sigma,\pi$ are elements of $n$th degree permutation group.

Here
\begin{equation}
(C_{\sigma\pi}) = C =Q^{-1},
\end{equation}
where
\begin{equation}
    Q = (Q_{\sigma\pi}),
\end{equation}
with the matrix elements
\begin{equation}
Q_{\sigma\pi} = 
%
\  .
\end{equation}

For example, for the $2$nd moment, the $Q$ matrix is 
\begin{equation}
Q = 
%
=d^2. 
\end{equation}

\paragraph{Second-order scar cumulants: $q=1$.}
Using the 1st moment Haar average \eqref{eqn:Haar_1st_moment}, we have
\begin{equation}
\begin{aligned}
\overline{O_{ai}O_{ib}} &= \overline{\bra{a}O\ket{i}\bra{i}O\ket{b}} \\
&= \int_{Haar}dU \bra{a}OU\ket{i}\bra{i}U^\dag O\ket{b} \\
&= \left(\frac{1}{d}\right) 
%
\\
&= \dfrac{1}{d}\bra{a}O^2\ket{b}
\sim e^{-S} k^{(2)}. 
\end{aligned}
\end{equation}
So here, we see that one index gives an $e^{-S}$ factor in the scar scaling form with $q=1$. 

\paragraph{Third-order scar cumulants: $q=2$.} 
Next, consider two thermal indices $q=2$. 
Using \eqref{eqn:Haar_2nd_moment}, we have
\begin{equation}
\begin{aligned}
\overline{O_{ai}O_{ij}O_{jb}} &= \overline{\bra{a}O\ket{i}\bra{i}O\ket{j}\bra{j}O\ket{b}} \\
&= \int_{Haar}dU \bra{a}OU\ket{i}\bra{i}U^\dag OU\ket{j}\bra{j}U^\dag O\ket{b} \\
&= 
\sum_{\sigma\pi}C_{\sigma\pi}\  
%
. 
\end{aligned}
\end{equation}
Here, the only choice to give non-zero contribution is $\pi=I$. So it becomes
\begin{equation}
\begin{aligned}
&=
C_{II}
%
\\
&= \dfrac{1}{d^2-1}\left(\bra{a}O^3\ket{b} + (-\dfrac{1}{d})\bra{a}O^2\ket{b}\Tr(O)\right) \\
&\simeq \dfrac{1}{d^2}\left(\bra{a}O^3\ket{b} - \bra{a}O^2\ket{b}\langle O\rangle\right) \\
&\sim e^{-2S} k_{ab}^{(3)}(O).  
\qquad\qquad \text{\small(here $k^{(3)}$ is a kind of the third order scar correlation function. )} 
\end{aligned}
\end{equation}
We see that, two thermal indices gives the $e^{-2S}$ factor. 
For higher order, using the general $n$th moment calculus, the leading order of Haar average over distinct $q$ thermal indices give rise to an $e^{-qS}$ factor, which is the scaling form of the SFETH ansatz we formulate in the maintext.

\bibliography{suppl_ref.bib}